\documentclass[journal]{IEEEtran}
\usepackage{ifpdf}
\usepackage{cite}
\usepackage[dvips]{graphicx}
\usepackage[cmex10]{amsmath}
\usepackage{amssymb}
\usepackage{amsmath}
\usepackage{mathrsfs}
\usepackage[lined,boxed,commentsnumbered, ruled]{algorithm2e}
\usepackage{array}
\usepackage{mdwmath}
\usepackage{eqparbox}
\usepackage[tight,footnotesize]{subfigure}
\usepackage{tikz}
\usetikzlibrary{arrows}
\usepackage{color}
\definecolor{red1}{cmyk}{0,1,.8,0}
\definecolor{blue1}{cmyk}{.9,.7,0,0}
\definecolor{blue2}{cmyk}{.93,.95,.2,.07}

\usepackage[caption=false,font=footnotesize]{subfig}

\newcommand{\bs}{\boldsymbol}

\begin{document}


\title{Scalable  Coordinated Beamforming for  Dense Wireless Cooperative
Networks
}
\author{\IEEEauthorblockN{Yuanming Shi, Jun Zhang, and Khaled B. Letaief,
\emph{Fellow, IEEE}}\\
\IEEEauthorblockA{Dept. of ECE, The Hong Kong University of Science and Technology\\
                           E-mail: \{yshiac, eejzhang, eekhaled\}@ust.hk}}

\maketitle
\IEEEpeerreviewmaketitle


\maketitle

\begin{abstract}
To meet the ever growing demand for both high throughput and uniform coverage in future wireless networks, dense network deployment will be ubiquitous, for which cooperation among the access points is critical. Considering the computational complexity of designing coordinated beamformers for dense networks, low-complexity and suboptimal precoding strategies are often adopted. However, it is not clear how much performance loss will be caused. To enable optimal coordinated beamforming, in this paper, we propose a framework to design a scalable beamforming algorithm based on the alternative direction method of multipliers (ADMM) method. Specifically, we first propose to apply the matrix stuffing technique to transform the original  optimization problem to an equivalent ADMM-compliant problem, which is much more efficient than the widely-used modeling framework CVX. We will then propose to use the ADMM algorithm, a.k.a. the operator splitting method, to solve the transformed ADMM-compliant problem efficiently. In particular, the subproblems of the ADMM algorithm at each iteration can be solved with closed-forms and in parallel. Simulation results show that the proposed techniques can result in significant computational efficiency compared to the state-of-the-art interior-point solvers. Furthermore, the simulation results demonstrate that the optimal coordinated beamforming can significantly improve the system performance compared to sub-optimal zero forcing beamforming.
\end{abstract}


\IEEEpeerreviewmaketitle

\section{Introduction}
The proliferation of smart mobile devices, coupled with new types of wireless applications, has led to an exponential growth of wireless and mobile data traffic.  In order to meet the data explosion, many advanced network architectures with dense deployment and coordination among the access points (APs) have been proposed for 5G cellular networks.  For instance, in network MIMO systems  \cite{Andrews_TIT2013}, all the APs are connected
through the backhaul links such that the channel state information (CSI) and user data can be shared among
the APs. By deploying a large number of APs, a large-scale full cooperative
network will be created.  In the recently proposed Cloud-RAN \cite{Yuanming_TWC2013,Yuanming_arXivOpt13}, all the
baseband signal processing is shifted to a single baseband unit (BBU)
pool with very powerful computational
capability.  The centralized signal processing is
performed at the BBU pool, to support large-scale cooperative transmission/reception among APs.

With the increasing demand of high capacity, it is critical to design efficient coordinated beamforming to take advantage of the full cooperation in dense networks. Considering the computational complexity, zero-forcing (ZF), regularized zero-forcing (RZF), and maximum ratio transmission (MRT) precoding are often adopted. However, such low-complexity strategies might significantly degrade system performance. The existing works on designing optimal coordinated beamforming
either apply the advanced off-the-shelf interior-point solvers like SeDuMi \cite{SeDuMi_1999using} or exploit the problem structures to design the custom solver, e.g., using the uplink-downlink duality theory \cite{WeiYu_WC10}. Although these algorithms have polynomial time complexity, such second-order methods
still are not efficient for the coordinated beamforming problems with a large
dimension. For practical implementation, the first-order method ADMM \cite{boyd2011distributed} has recently been widely used to solve large-scale optimization problems to modest accuracy within reasonable time. However, most existing works (e.g., \cite{shen2012distributed}) on applying the ADMM algorithms still need to solve convex problems (e.g., semidefinite programming (SDP) problem) for the subproblems at each iteration, though they enjoy the distributed implementation property. Therefore, in order to improve the efficiency of the ADMM algorithm, we need to reduce the computational complexity for solving the subproblems in the ADMM algorithm.  

In this paper, we propose a framework to design a scalable optimization algorithm based on the ADMM method to solve the coordinated beamforming problems in dense networks, with the number of cooperating APs as high as 100. By introducing a new variable for each subexpression in the original coordinated beamforming problem based on the \emph{Smith form} reformulation \cite{boyd_code2013}, we transform the original problem into the equivalent ADMM-compliant problem. For any fixed size networks (i.e., given the number of APs and MUs), the structure of the ADMM-compliant problem is fixed. Therefore, we can first generate and store the structure of the ADMM-compliant problem for any network  with a  fixed size, which can be done offline. For any particular network realization, we only need to copy the parameters of the original problem to the ADMM-compliant problem. A similar idea was presented in \cite{boyd_code2013} for embedded systems, called the \emph{matrix stuffing} technique. Compared to the modeling framework CVX \cite{cvx}, this method only needs to copy the memory and can be much more efficient. We shall propose to use the ADMM method to solve the transformed ADMM-compliant problem. With the special structure of the transformed ADMM-compliant problem, we can solve the subproblems of the corresponding ADMM algorithm at each iteration with closed-forms and in parallel. Simulation results will demonstrate  speedups of
several orders of magnitude of the proposed techniques
for large-scale coordinated beamforming over the state-of-the-art interior-point solvers. Furthermore, simulation
results will show that the optimal coordinated beamforming outperforms zero forcing beamforming significantly.

\section{Coordinated Beamforming for Maximizing the minimum network-wide achievable rate}
Consider a cellular network with $L$ access points (APs), where the $l$-th AP is equipped with $N_{l}$ antennas, and with $K$ single-antenna mobile users (MUs). We mainly consider the full cooperative networks, though the algorithms proposed in this paper can be easily extended to other kinds of cooperation strategies. The critical question is how to take full advantage of the cooperation with scalable and efficient coordinated beamforming algorithms instead of simply adopting the suboptimal strategies such as zero-forcing beamforming. In this paper, we mainly focus on maximizing the minimum network-wide achievable rate. This problem is formulated as the following max-min fairness
optimization problem with per-AP transmit power constraints:
\begin{eqnarray}
\label{mm}
\mathop {\rm{maximize~}}_{{\bf{v}}}&&\left(\min_{1\le k\le K}\omega_{k}\log_2(1+{{\Gamma}}_{k}({\bf{v}};
{\bf{h}}_k))\right)\nonumber\\
{\rm{subject~to~}}&&  \sum_{k=1}^K\|{\bf{v}}_{lk}\|_2^2\le P_l, \forall l,
\end{eqnarray}
where $\omega_k>0$ is the weight for MU $k$ and $\Gamma_{k}({\bf{v}}; {\bf{h}}_k)$ is the achievable SINR at MU $k$ using the single user detection 
\setlength\arraycolsep{1pt}
\begin{eqnarray}
{{\Gamma}}_{k}({\bf{v}}; {\bf{h}}_k)={{|{\bf{h}}_k^{\sf{H}}{\bf{v}}_k|^2}\over{\sum_{i\ne
k}|{\bf{h}}_{k}^{\sf{H}}{\bf{v}}_i|^2}+\sigma_k^2}, \forall k,
\end{eqnarray}
where ${\bf{h}}_{k}= [{\bf{h}}_{k1}^{T},{\bf{h}}_{k2}^{T},\dots,{\bf{h}}_{kL}^{T}]^{T}\in\mathbb{C}^{N}$
with ${\bf{h}}_{kl}\in\mathbb{C}^{N_l}$ as the channel coefficient vector between AP $l$ and MU $k$ and $N=\sum_{l=1}^{L}N_{l}$, and ${\bf{v}}_{k}=[{\bf{v}}_{1k}^{T},
{\bf{v}}_{2k}^{T},\dots, {\bf{v}}_{Lk}^{T}]^{T}\in\mathbb{C}^{N}$ with ${\bf{v}}_{lk}\in\mathbb{C}^{N_l}$ as the beamforming vector at AP $l$ for MU $k$, and $P_l$ is the maximum transmit power at AP $l$.

Problem (\ref{mm}) can be solved by
a bi-section method \cite{boyd2004convex}. Specifically, given a threshold $\gamma$, we need to
solve the following optimization
problem
\begin{eqnarray}
\label{feasible}
\mathscr{P}:
\mathop {\rm{minimize~}} && \|{{\bf{v}}}\|_2\nonumber\\
{\rm{subject~to~}}&&  \omega_{k}\log_2(1+{{\Gamma}}_{k}({\bf{v}};
{\bf{h}}_k))\ge\gamma, \forall
k\nonumber\\
&&\sum_{k=1}^K\|{\bf{v}}_{lk}\|_2^2\le P_l, \forall l,
\end{eqnarray}
which can be reformulated as an SOCP problem \cite{Yuanming_TWC2013,WeiYu_WC10}.  Therefore, the max-min fairness optimization algorithm is presented as Algorithm 1.
\begin{algorithm}
\caption{Max-min Fairness Optimization Algorithm }
\begin{itemize}
\item Initialize the $\gamma_{\textrm{low}}=0$ and $\gamma_{\textrm{up}}=\gamma_{\textrm{max}}$;\\
\item Repeat
\begin{enumerate}
\item Set $\gamma\leftarrow (\gamma_{\textrm{low}}+\gamma_{\textrm{up}})/2$;
\item Solve problem $\mathscr{P}$ with the given $\gamma$; if it is feasible,\\
set $\gamma_{\textrm{low}}\leftarrow\gamma$; otherwise, set $\gamma_{\textrm{up}}\leftarrow\gamma$;
\end{enumerate}
\item Until $r_{\textrm{up}}-r_{\textrm{low}}<\epsilon$, where
$\epsilon$ represents the accuracy requirement.
\end{itemize}  
\end{algorithm}

\subsection{Problem Analysis}
The main computational complexity of the optimal coordinated beamforming  is solving the convex optimization problem $\mathscr{P}$ with different SINR thresholds. In order to solve the problem $\mathscr{P}$, one might use the advanced off-the-shelf interior-point solvers like SeDuMi \cite{SeDuMi_1999using}. But such second-order method is unable to handle problems with a large dimension, as it has the computational complexity $\mathcal{O}(N^{3.5}K^{3.5})$. Moreover, in order to use the standard interior-point solvers, one needs to transform the problem $\mathscr{P}$ into a standard form. The modeling framework CVX \cite{cvx} can carry out a sequence of equivalence transformations to yield a problem that can be handled by a standard interior-point solver. But this procedure is not efficient, as for each problem instance (i.e., given the problem parameters), CVX will check the convexity of the problem $\mathscr{P}$ and then transform it to a standard form using the graph implementation \cite{boyd2008graph} to represent the convex functions. In order to solve the large-scale optimization problem $\mathscr{P}$, in this paper, we propose a framework based on the ADMM method.

\subsubsection{ADMM Algorithm} 
For practical implementation of the coordinated beamforming, we propose to use the first-order method, i.e., the ADMM algorithm, to solve the large-scale optimization problem $\mathscr{P}$. Consider the following convex optimization problem of the form
\begin{eqnarray}
\label{admmo}
\mathop {\rm{maximize~}}_{{\bf{x}}, {\bf{y}}}&&f({\bf{x}})+g({\bf{y}})\nonumber\\
{\rm{subject~to~}}&&  {\bf{x}}-{\bf{y}}={\bf{0}},
\end{eqnarray}
where $f({\bf{x}})$ and $g({\bf{y}})$ are convex functions. The ADMM algorithm solves problem (\ref{admmo}) iteratively by the following three steps at each iteration $k$:
\begin{eqnarray}
\label{admm1}
{\bf{x}}^{[k+1]}&=&\arg\min_{\bf{x}} (f({\bf{x}})+(\rho/2)\|{\bf{x}}-{\bf{y}}^{[k]}+{\boldsymbol{\lambda}}^{[k]}\|_2^2),\\
\label{admm2}
{\bf{y}}^{[k+1]}&=&\arg\min_{\bf{y}} (g({\bf{y}})+(\rho/2)\|{\bf{x}}^{[k+1]}-{\bf{y}}+{\boldsymbol{\lambda}}^{[k]}\|_2^2),\\
{\boldsymbol{\lambda}}^{[k+1]}&=&{\boldsymbol{\lambda}}^{[k]}+{\bf{x}}^{[k+1]}-{\bf{y}}^{[k+1]},
\end{eqnarray} 
where $\rho>0$ is a step size parameter and $\boldsymbol{\lambda}$ is the dual variable associated with the constraint ${\bf{x}}-{\bf{y}}={\bf{0}}$. Under some very mild conditions \cite[Section 3.2]{boyd2011distributed} (e.g., the functions $f$ and $g$ are closed, proper, and convex),  the solutions obtained from the ADMM algorithm converge to a global optimal solution, i.e., $f({\bf{x}}^{[k]})+g({\bf{y}}^{[k]})$ converges to the optimal value, ${\boldsymbol{\boldsymbol{\lambda}}}^{[k]}$ converges to an optimal dual variable, and ${\bf{x}}^{[k]}-{\bf{y}}^{[k]}$ converges to zero.

The main advantage of the ADMM algorithm is that the subproblems (\ref{admm1}) and (\ref{admm2}) can be solved with closed-form and in parallel
in many applications, e.g., in machine learning and high dimensional statistics. However, for the coordinated beamforming problem $\mathscr{P}$,  it is not clear how to reformulate the problem $\mathscr{P}$ such that the subproblems of the corresponding ADMM algorithm can be solved with closed-forms and in parallel. Although there are some works applying the ADMM algorithm to solve the coordinated beamforming problems \cite{shen2012distributed} with distributed implementation, they still need to solve SDP problems for the subproblems of the ADMM algorithm at each iteration. This is not efficient in terms of computational complexity.

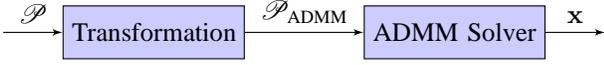
\begin{figure}[!t]
\centering
\tikzstyle{int}=[draw, fill=blue!20, minimum size=2em]
\tikzstyle{init} = [pin edge={to-,thin,black}]
\begin{tikzpicture}
    [node distance=4cm,auto,>=latex']
    \node [int, ] (a) {Transformation};
    \node (b) [left of=a,node distance=2cm, coordinate] {a};
    \node [int, ] (c) [right of=a] {ADMM Solver};
    \node [coordinate] (end) [right of=c, node distance=2.0cm]{};
    \path[->] (b) edge node {$\mathscr{P}$} (a);
    \path[->] (a) edge node {$\mathscr{P}_{\textrm{ADMM}}$} (c);
    \draw[->] (c) edge node {$\bf{x}$} (end) ;
\end{tikzpicture}
\caption{The framework of the large-scale optimization algorithm}
\label{dcp}
\end{figure}

To apply the ADMM algorithm to solve the large-scale optimization problem $\mathscr{P}$ efficiently, in this paper, we propose a framework as shown in Fig. {\ref{dcp}}. It will first transform the original problem $\mathscr{P}$ to the equivalent ADMM-compliant form $\mathscr{P}_{\textrm{ADMM}}$ as follows based  the Smith form reformulation \cite{boyd_code2013}:
\begin{eqnarray}
\mathscr{P}_{\textrm{ADMM}}:
\mathop {\rm{minimize~}}_{{\bs{\nu}}, {\bs{\mu}}}&&{\bf{c}}^{T}{\bs{\nu}}\nonumber\\
{\rm{subject~to~}}&& {\bf{A}}{\bs{\nu}}+{\bs{\mu}}={\bf{b}}\nonumber\\
&& ({\bs{\nu}}, {\bs{\mu}})\in\mathbb{R}^{n}\times\mathcal{V}.
\end{eqnarray}
where ${\bs{\nu}}\in\mathbb{R}^n$ and ${\bs{\mu}}\in\mathbb{R}^{m}$ are the
optimization variables, $\mathcal{V}=\{0\}^{r}\times\mathcal{Q}^{m_1}\times\cdots\times\mathcal{Q}^{m_q}$
with ${\mathcal{Q}}^{p}=\{(t, {\bf{x}})\in\mathbb{R}\times\mathbb{R}^{p-1}|
\|{\bf{x}}\|\le t\}$ as the second-order cone of dimension $p$, and $\mathcal{Q}^{1}$ is defined as the cone of nonnegative reals, i.e., $\mathbb{R}_{+}$.
Here, each $\mathcal{Q}^i$
has dimension $m_i$ such that $(r+\sum_{i=1}^q m_i)=m$, ${\bf{A}}\in\mathbb{R}^{m\times
n}$, ${\bf{b}}\in\mathbb{R}^{m}$, ${\bf{c}}\in\mathbb{R}^{n}$. The idea of Smith form reformulation is simple and we only need to introduce a new variable for each subexpression in the original problem $\mathscr{P}$. We then propose to use the ADMM algorithm to solve the homogeneous self-dual embedding \cite{ye1994nl} of the problem$\mathscr{P}_{\textrm{ADMM}}$. As a result, each subproblem in the corresponding ADMM algorithm can be solved with closed-forms and in parallel. This will be much more efficient than the previous works on applying the ADMM algorithms to design the coordinated beamformers \cite{shen2012distributed}.

\section{Matrix stuffing for Fast Transformation}
\label{mats}
In this section, we propose a framework to transform the original
problem $\mathscr{P}$ to the equivalent ADMM-compliant problem $\mathscr{P}_{\textrm{ADMM}}$, i.e., for the first stage of Fig. 1. Although CVX can perform this procedure automatically, it needs to carry out a sequence of equivalence transformations. This is not efficient, especially with a large number of constraints in a dense network. 

Instead, in this paper, we propose to transform the original problem
$\mathscr{P}$ using the {Smith form} reformulation, which involves introducing a new variable for each subexpression in the functions of problem $\mathscr{P}$. This transformation has two advantages: first, the parameters in the original problem $\mathscr{P}$ only appear in the affine functions of the transformed ADMM-compliant problem $\mathscr{P}_{\textrm{ADMM}}$, which is suitable for matrix stuffing as will be shown in Section {\ref{ms}}; second, the cone in $\mathscr{P}_{\textrm{ADMM}}$ is a Cartesian product of the standard cones, thus the subproblems (\ref{admm1}) and (\ref{admm2}) of the corresponding ADMM algorithm can be solved with closed-forms and in parallel. We first deal with the real-field case,  the extension to the complex-field case will be discussed in Section \ref{complex}.

\subsection{ADMM-Compliant Form Reformulation for Problem $\mathscr{P}$}
In this subsection, we reformulate the real-field problem $\mathscr{P}$ by introducing a new variable for each subexpression in $\mathscr{P}$, resulting in the equivalent ADMM-compliant form $\mathscr{P}_{\textrm{ADMM}}$.

Specifically, for the objective function $f({\bf{v}})=\|{\bf{v}}\|_2$, by introducing the new variables $x_0$ and ${\bf{x}}_1$, minimizing $f({\bf{v}})$ is equivalent to minimizing $x_0$ with the following constraints
\begin{eqnarray}
\mathcal{G}_1:\left\{\begin{array}{l}
(x_0, {\bf{x}}_1)\in\mathcal{Q}^{M+1}\\
{\bf{x}}_1={\bf{v}}\in\mathbb{R}^{M},
\end{array}\right.
\end{eqnarray}
where $M=KN$. With variable $[x_0; {\bf{v}}]$ and the same order of equations as in $\mathcal{G}_1$, $\mathcal{G}_1$ can be rewritten as 
\begin{eqnarray}
{\bf{M}}[{{x}}_0; {\bf{v}}]+{\bs{\mu}}_1={\bf{m}},
\end{eqnarray}
with ${\bf{M}}={\rm{blkdiag}}\{-1,-{\bf{I}}_M\}\in\mathbb{R}^{(M+1)\times(M+1)}$
and ${\bf{m}}=[0,{\bf{0}}_M^T]^{T}$ and ${\bs{\mu}}_1\in\mathcal{Q}^{M+1}$.

For the per-AP transmit power constraint $\|\tilde{\bf{v}}_l\|_2\le \sqrt{P_l}$
with $\tilde{\bf{v}}_l=[{\bf{v}}_{l1}^{T},\dots, {\bf{v}}_{lK}^{T}]^{T}\in\mathbb{R}^{KN_l}$,
we first have the following
equivalent expressions 
\begin{eqnarray}
\|\tilde{\bf{v}}_l\|_2\le\sqrt{P_l}\Longrightarrow\|{\bf{D}}_l{\bf{v}}\|_2\le
\sqrt{P_l},
\end{eqnarray}
where ${\bf{D}}_{l}={\rm{blkdiag}}\{{\bf{D}}_l^1,\dots,{\bf{D}}_l^K\}\in\mathbb{R}^{KN_l\times
M}$ with ${\bf{D}}_{l}^k=\left[{\bf{0}}_{N_{l}\times\sum_{i=1}^{l-1}N_i},{\bf{I}}_{N_l\times
N_l}, {\bf{0}}_{{N_l}\times
\sum_{i=l+1}^{L}N_i}\right]\in\mathbb{R}^{N_{l}\times N}$ such that ${\bf{D}}_l{\bf{v}}\triangleq\tilde{\bf{v}}_l$. By introducing the new variables $y_{0}^l$ and ${\bf{y}}_1^l$, we have the following equivalent formulation for the constraint $\|{\bf{D}}_l{\bf{v}}\|_2\le
\sqrt{P_l}$, i.e.,
\begin{eqnarray}
\mathcal{G}_2(l):\left\{\begin{array}{l}
 (y_0^l, {\bf{y}}_1^{l})\in\mathcal{Q}^{KN_l+1}\\
y_0^l=\sqrt{P_l}\in\mathbb{R}\\
{\bf{y}}_1^l={\bf{D}}_l{\bf{v}}\in\mathbb{R}^{KN_l}.
\end{array}\right.
\end{eqnarray}
With variable $[y_0^l; {\bf{v}}]$ and the same order
of equations as in $\mathcal{G}_2(l)$, $\mathcal{G}_2(l)$ can be rewritten as
\begin{eqnarray}
{\bf{P}}^l[{{y}}_0^{l}; {\bf{v}}]+{\bs{\mu}}_2^l={\bf{p}}^l,
\end{eqnarray}
where 
\begin{eqnarray}
\!\!\!\!\!{\mathbf{P}}^l =
\left[ \begin{array}{rc}
1&  \\
\hline
-1&   \\
&   -{\bf{D}}_l
 
\end{array} \right]\in\mathbb{R}^{(KN_l+2)\times(M+1)}, 
{{\bf{p}}}^l=\left[ \begin{array}{c}
\sqrt{P_l} \\
\hline
0\\
{\bf{0}}_{KN_l}
\end{array} \right],
\end{eqnarray}
and ${\bs{\mu}}_2^l\in\mathcal{Q}^{1}\times\mathcal{Q}^{KN_l+1}$.

For the QoS constraint of MU $k$, let $\theta_k=2^{\gamma/\omega_k}-1$,
we have the following equivalent expressions
\begin{eqnarray}
\label{qs1}
\!\!\!\!\!\!{{|{\bf{h}}_k^{{T}}{\bf{v}}_{k}|^2}\over{{\sum\nolimits_{i\ne
k}|{\bf{h}}_k^{{T}}{\bf{v}}_{i}|^2}}+\sigma_{k}^2}\ge \theta_{k}\Longrightarrow\|{\bf{C}}_k{\bf{v}}+{\bf{g}}_k\|_2\le
\beta_{k}{\bf{r}}_k^{T}{\bf{v}},
\end{eqnarray}
where $\beta_k=\sqrt{1+1/\theta_k}\in\mathbb{R}$, ${\bf{g}}_k=[{\bf{0}}_K^T,
\sigma_k]^T\in\mathbb{R}^{K+1}$, ${\bf{C}}_k\in\mathbb{R}^{(K+1)\times
M}$ is
given by
\begin{eqnarray}
\mathbf{C}_k =
\left[ \begin{array}{ccc}
{\bf{h}}_k^{T} & & \\
&\ddots &\\
& & {\bf{h}}_k^{T} \\
 \hline\\[-3.7mm]
&{\bf{0}}_{M}^{T} &
\end{array} \right]\in\mathbb{R}^{(K+1)\times M},
\end{eqnarray}
and ${\bf{r}}_k=\left[{\bf{0}}_{(k-1)\sum_{l=1}^LN_l}^{T},{\bf{h}}_k^{T},
{\bf{0}}_{(K-k)\sum_{l=1}^LN_l}^{T}\right]^{T}\in\mathbb{R}^M$. By introducing the new variables $t_0^k$ and ${\bf{t}}_1^k$, we have the following equivalent formulation for $\|{\bf{C}}_k{\bf{v}}+{\bf{g}}_k\|_2\le
\beta_{k}{\bf{r}}_k^{T}{\bf{v}}$, i.e., 
\begin{eqnarray}
\mathcal{G}_3(k):\left\{\begin{array}{l}
(t_0^k, {\bf{t}}_1^k)\in\mathcal{Q}^{K+1}\\
t_0^k=\beta_{k}{\bf{r}}_k^{T}{\bf{v}}\in\mathbb{R}\\
{\bf{t}}_1^k={\bf{t}}_2^k+{\bf{t}}_3^k\in\mathbb{R}^{K+1}\\
{\bf{t}}_2^k={\bf{C}}_k{\bf{v}}\in\mathbb{R}^{K+1}\\
{\bf{t}}_3^k={\bf{g}}_k\in\mathbb{R}^{K+1}.
\end{array}\right.
\end{eqnarray}
As a result, with variable $[t_0^k; {\bf{v}}]$ and the same order
of equations as in $\mathcal{G}_3(k)$, $\mathcal{G}_3(k)$ can be rewritten as 
 \begin{eqnarray}
{\bf{Q}}^k[t_{0}^k;{\bf{v}}]+{\bs{\mu}}_3^k={\bf{q}}^k,
\end{eqnarray}
where 
\begin{eqnarray}
{\mathbf{Q}}^k =
\left[ \begin{array}{rc}
1& -\beta_{k}{\bf{r}}_k^{T} \\
\hline
-1&   \\
&   -{\bf{C}}_k
 
\end{array} \right]\in\mathbb{R}^{(K+3)\times(M+1)}, 
{{\bf{q}}}^k=\left[ \begin{array}{c}
0 \\
\hline
0\\
{\bf{g}}_k
\end{array} \right],
\end{eqnarray}
and ${\bs{\mu}}_3^k\in\mathcal{Q}^{1}\times \mathcal{Q}^{K+2}$. 

Now we arrive at the following ADMM-compliant problem $\mathscr{P}_{\textrm{ADMM}}$ 
with the
optimization variables given by ${\bs{\nu}}=[x_0; y_0^{1};\dots; y_0^{L};
t_0^1; \dots, t_0^K; {\bf{v}}]\in\mathbb{R}^{n}$
and ${\bf{c}}=[1; {\bf{0}}_{n-1}]$. The ADMM-compliant problem $\mathscr{P}_{\textrm{ADMM}}$
structure is characterized by the following data
\begin{eqnarray}
\label{ds1}
n&=&1+L+K+M,\\
m&=&(L+K)\!+\!(M+1)\!+\!\sum_{l=1}^{L}(KN_l+1)\!+\!K(K+2),\\
\mathcal{V}&=& \underbrace{\mathcal{Q}^{1}\times\cdots\times\mathcal{Q}^1}_{L+K}\times\mathcal{Q}^{M+1}\times
\underbrace{\mathcal{Q}^{KN_1+1}\times\dots\times\mathcal{Q}^{KN_L+1}}_{L}\nonumber\\
&&\times\underbrace{\mathcal{Q}^{K+2}\times\dots\times\mathcal{Q}^{K+2}}_{K},
\end{eqnarray}
where $\mathcal{V}$ is the Cartesian product of $2(L+K)+1$ closed convex ones, and
 ${\bf{A}}$ and $\bf{b}$ are given as
follows: 
\begin{eqnarray}
\label{ds2}
\!\!\!\!\!\!\!\!\!\!{\mathbf{A}} =
\left[ \begin{array}{ccccccc|c}
&1&&&&&\\
&&\ddots&&&&\\
&&&1&&&\\
\hline
&&&&1&&&-\beta_{1}{\bf{r}}_1^{T}\\
&&&&&\ddots&&\vdots\\
&&&&&&1&-\beta_{K}{\bf{r}}_K^{T}\\
\hline
-1&&&&&&&\\
&&&&&&&-{\bf{I}}_B\\
\hline
&-1&&&&&&\\
&&&&&&&-{\bf{D}}_1\\
\hline
&&&\vdots&&&&\vdots\\
\hline
&&&-1&&&&\\
&&&&&&&-{\bf{D}}_L\\
\hline
&&&&-1&&&\\
&&&&&&&-{\bf{C}}_1\\
\hline
&&&\vdots&&&&\vdots\\
\hline
&&&&&&-1&\\
&&&&&&&-{\bf{C}}_K\\
\end{array} \right], 
{\bf{b}}=\left[ \begin{array}{c}
\sqrt{P_1}\\
\vdots\\
\sqrt{P_L}\\
\hline
0\\
\vdots\\
0\\
\hline
{\bf{0}}_{B}\\
\hline
0\\
{\bf{0}}_{KN_1}\\
\hline
\vdots\\
\hline
0\\
{\bf{0}}_{KN_1}\\
\hline
0\\
{\bf{g}}_1\\
\hline
\vdots\\
\hline
0\\
{\bf{g}}_K
\end{array}\right],
\end{eqnarray}

\subsection{Matrix Stuffing for Transformation}
\label{ms}
Given the network size, i.e., the number of APs and MUs, the structure of the ADMM-compliant problem $\mathscr{P}_{\textrm{ADMM}}$ is fixed. Therefore, we can first generate and store the problem structure of $\mathscr{P}_{\textrm{ADMM}}$ for a fixed-size network, i.e., the structure of ${\bf{A}}$, $\bf{b}$, $\bf{c}$, and the descriptions of $\mathcal{V}$. This procedure can be done offline as the number of APs and MUs will keep constant  for a long period. Then for any specific network realization, we only need to
copy the parameters of the original problem  to the corresponding data in $\mathscr{P}_{\textrm{ADMM}}$. Specifically, we only need to copy the parameters of the maximum transmit power $P_{l}$'s to the data of the ADMM-compliant problem, i.e.,  $\sqrt{P_l}$'s in $\bf{b}$, copy the parameters of the SINR thresholds $\gamma$ to the data of the ADMM-compliant problem, i.e., $\beta_k$'s in ${\bf{A}}$, and copy the parameters of the channel realizations ${\bf{h}}_k$'s to the data of the ADMM-compliant problem, i.e., ${\bf{r}}_k$'s and ${\bf{C}}_k$'s in $\bf{A}$. As we only need to perform copying the memory for the transformation, this procedure can be very efficient compared to the state-of-the-art modeling framework CVX.

\subsection{Extension to the Complex Case}
\label{complex}
In this subsection, we present how to extend the real-field problem to the complex-field problem. For ${\bf{h}}_k\in\mathbb{C}^{N},
{\bf{v}}_{i}\in\mathbb{C}^{N}$, we have
\begin{eqnarray}
{\bf{h}}_k^{\sf{H}}{\bf{v}}_i\Longrightarrow
{\underbrace{\left[ \begin{array}{cc}
\mathfrak{R}({\bf{h}}_k)\ & -\mathfrak{J}({\bf{h}}_k) \\
\mathfrak{J}({\bf{h}}_k)\ & \mathfrak{R}({\bf{h}}_k) 
\end{array} \right]}_{\tilde{\bf{h}}_k}}^{T}\underbrace{\left[ \begin{array}{c}
\mathfrak{R}({\bf{v}}_i)\ \\
\mathfrak{J}({\bf{v}}_i) 
\end{array} \right]}_{\tilde{\bf{v}}_i},
\end{eqnarray} 
where $\tilde{\bf{h}}_k\in\mathbb{R}^{2N\times
2N}$ and $\tilde{\bf{v}}_i\in\mathbb{R}^{2N}$. Therefore, the complex-field problem can be changed into the real-field problem by
the transformations: ${\bf{h}}_k\Rightarrow\tilde{\bf{h}}_k~{\textrm{and}}~{\bf{v}}_i\Rightarrow\tilde{\bf{v}}_i$.

\section{The ADMM Algorithm For Large-Scale Coordinated Beamforming}
\label{os}
In this section, we propose
to use the ADMM algorithm \cite{boyd2011distributed}, a.k.a. the operator splitting method \cite{Boyd_arXiv2013}, to solve the ADMM-compliant problem $\mathscr{P}_{\textrm{ADMM}}$. Specifically,
we first introduce the homogeneous self-dual embedding \cite{ye1994nl} for the ADMM-compliant problem $\mathscr{P}_{\textrm{ADMM}}$. Then we use the ADMM algorithm \cite{boyd2011distributed,Boyd_arXiv2013} instead of the interior-point method to solve the embeddings with a large dimension. This approach can solve large-scale optimization problems to modest accuracy very efficiently and can also be parallelizable  across multiple processors. 

\subsection{Homogeneous Self-Dual Embedding}
The dual problem of $\mathscr{P}_{\textrm{ADMM}}$ is given by
\begin{eqnarray}
\mathscr{D}_{\textrm{ADMM}}:
\mathop {\rm{minimize}}_{{\boldsymbol{\eta}}, {\boldsymbol{\lambda}}}&&-{\bf{b}}^{T}{\boldsymbol{\eta}}\nonumber\\
{\rm{subject~to}}&& -{\bf{A}}^{T}{\boldsymbol{\eta}}+{\boldsymbol{\lambda}}={\bf{c}}\nonumber\\
&& ({\boldsymbol{\lambda}}, {\boldsymbol{\eta}})\in\{0\}^{n}\times\mathcal{V}^{*},
\end{eqnarray}
where $\boldsymbol{\lambda}\in\mathbb{C}^{n}$ and $\boldsymbol{\eta}\in\mathbb{C}^m$
are the dual variables, $\mathcal{V}^*$ is the dual cone of the non-empty
closed convex cone $\mathcal{V}$ and $\{0\}^n$ is the dual cone of $\mathbb{R}^n$. The Karush-Kuhn-Tucker (KKT) conditions are necessary and sufficient for optimality when strong duality holds. Specifically, when\begin{eqnarray}
{\bf{A}}{\bs{\nu}}^{\star}+{\bs{\mu}}^{\star}&=&{\bf{b}}, {\bs{\mu}}^{\star}\in\mathcal{V}, {\bf{c}}^{T}{\bs{\nu}}^{\star}+{\bf{b}}^{T}{\boldsymbol{\eta}}^{\star}={\bf{0}}, \nonumber\\ {\bf{A}}^{T}{\boldsymbol{\eta}}^{\star}+{\bf{c}}&=& {\boldsymbol{\lambda}}^{\star}, {\boldsymbol{\lambda}}^{\star}={\bf{0}}, {\boldsymbol{\eta}}^{\star}\in\mathcal{V}^{*},
\end{eqnarray}
$({\boldsymbol{\nu}}^{\star},
{\boldsymbol{\mu}}^{\star}, {\boldsymbol{\lambda}}^{\star}, {\boldsymbol{\eta}}^{\star})$
satisfies the KKT conditions and is primal-dual optimal. By introducing two new nonnegative variables ${{\tau}}$ and ${{\kappa}}$, the original primal-dual problems $\mathscr{P}_{\textrm{ADMM}}$ and $\mathscr{D}_{\textrm{ADMM}}$ can be converted into a single feasibility
problem by embedding the Karush-Kuhn-Tucker (KKT) conditions into the following
single system of equations, i.e., the homogeneous self-dual embedding \cite{ye1994nl}: 
\begin{eqnarray}
\label{emb}
\underbrace{\left[ \begin{array}{c}
{\bs{\lambda}} \\
{\boldsymbol{\mu}} \\
{{\kappa}}
\end{array} \right]}_{\bf{y}}=\underbrace{\left[ \begin{array}{ccc}
{\bf{0}} & {\bf{A}}^{T} & {\bf{c}} \\
-{\bf{A}} & {\bf{0}} & {\bf{b}} \\
-{\bf{c}}^{T} & -{\bf{b}}^{T} & {\bf{0}}
\end{array} \right]}_{\bf{Q}}
\underbrace{\left[ \begin{array}{c}
{\bs{\nu}} \\
{\boldsymbol{\eta}} \\
{{\tau}}
\end{array} \right]}_{\bf{x}}, 
\end{eqnarray}
where $({\bf{x}}, {\bf{y}})\in \mathcal{C}\times \mathcal{C}^{*}$ with $\mathcal{C}=\mathbb{R}^n\times \mathcal{V}^*\times 
\mathbb{R}_+$ and $\mathcal{C}^{*}=\{0\}^n\times\mathcal{V}\times\mathbb{R}_{+}$.

\subsection{ADMM Algorithm for the Homogeneous Self-Dual Embedding}
To apply the 
 ADMM algorithm, we first transform the embedding system (\ref{emb}) to the following ADMM form
\begin{eqnarray}
\label{admm}
\mathscr{P}_{\textrm{Emb}}: 
\mathop {\rm{minimize}}_{{\bf{x}}, \tilde {\bf{x}},{\bf{y}},  \tilde{\bf{y}}}&& I_{\mathcal{C}\times\mathcal{C}^{*}}({\bf{x}}, {\bf{y}})+I_{{\bf{Q}}\tilde{\bf{x}}=\tilde{\bf{y}}}(\tilde{\bf{x}}, \tilde{\bf{y}})\nonumber\\
{\rm{subject~to}}&& ({\bf{x}}, {\bf{y}})=(\tilde{\bf{x}}, \tilde{\bf{y}}),
\end{eqnarray}
where $I_{\mathcal{S}}$ is the indicator function of the set $\mathcal{S}$. Applying the ADMM algorithm to the problem $\mathscr{P}_{\textrm{Emb}}$, the final algorithm is shown as follows \cite{Boyd_arXiv2013}:
\begin{eqnarray}
\tilde{\bf{x}}^{[i+1]}&=&({\bf{I}}+{\bf{Q}})^{-1}({\bf{x}}^{[i]}+{\bf{y}}^{[i]})\\
\label{prox}
{\bf{x}}^{[i+1]}&=&\Pi_{\mathcal{C}}(\tilde{\bf{x}}^{[i+1]}-{\bf{y}}^{[i]})\\
{\bf{y}}^{[i+1]}&=&{\bf{y}}^{[i]}-\tilde{\bf{x}}^{[i+1]}+{\bf{x}}^{[i+1]},
\end{eqnarray}
where $\Pi_{\mathcal{S}}({\bf{x}})$ denotes the Euclidean projection of $\bf{x}$ onto the set $\mathcal{S}$. The first step is performing projection onto a subspace, i.e., solving a linear system with the coefficient matrix ${\bf{I}}+{\bf{Q}}$. Some efficient algorithms can be found in \cite{Boyd_arXiv2013}. The last step is computationally trivial. The second step is performing projection onto the cone $\mathcal{C}$. As $\mathcal{C}$ is the Cartesian produce of the cones $\mathcal{C}_i$, we can project onto $\mathcal{C}$ by projecting onto $\mathcal{C}_i$ separately and in parallel. Furthermore, the projection onto  cones can be done with closed-forms and can be very efficient. For instance, for $\mathcal{C}_i=\mathbb{R}_{+}$, we have that \cite[Section 6.3]{Boyd_2013proximal}
\begin{eqnarray}
\Pi_{\mathcal{C}_i}({\boldsymbol{\omega}})={\boldsymbol{\omega}}_+,
\end{eqnarray}
where the nonnegative part operator $(\cdot)_+$ is taken elementwise. For the second-order cone ${\mathcal{C}}_i=\{(t, {\bf{x}})\in\mathbb{R}\times\mathbb{R}^{p-1}|
\|{\bf{x}}\|\le t\}$, we have that \cite[Section 6.3]{Boyd_2013proximal}
\begin{eqnarray}
\Pi_{\mathcal{C}_i}({\boldsymbol{\omega}},\tau)=\left\{\begin{array}{l}
0, \|{\boldsymbol{\omega}}\|_2\le -\tau\\
({\boldsymbol{\omega}},\tau), \|{\boldsymbol{\omega}}\|_2\le \tau\\
(1/2)(1+\tau/\|{\boldsymbol{\omega}}\|_2)({\boldsymbol{\omega}},\|{\boldsymbol{\omega}}\|_2), \|{\boldsymbol{\omega}}\|_2\ge \tau.
\end{array}\right.
\end{eqnarray}
The details on the convergence and termination criteria of this ADMM algorithm can be found in \cite{Boyd_arXiv2013}.

\section{Simulation Results}
\label{simu}
In this section, we simulate the large-scale optimization framework consists of matrix stuffing and the ADMM solver for the large-scale coordinated beamforming problems. We consider
the following channel model for the link between the $k$-th MU and the
$l$-th AP:
\begin{eqnarray}
{\bf{h}}_{kl}=\underbrace{10^{-L(d_{kl})/20}\sqrt{\varphi_{kl}s_{kl}}}_{D_{kl}}{\bf{f}}_{kl},
\forall k,l,
\end{eqnarray}
where $L(d_{kl})$ is the path-loss at distance $d_{kl}$, as given in \cite[Table
I]{Yuanming_TWC2013}, $s_{kl}$ is the shadowing coefficient, $\varphi_{kl}$ is
the antenna gain and ${\bf{f}}_{kl}$ is the small-scale fading coefficient.
We use the standard cellular network parameters as showed in \cite[Table I]{Yuanming_TWC2013}. All the simulations are carried out on a personal computer with 2.3 GHz Intel Core i7 processor and 8 GB of RAM running  OS X 10.9.2. 

\subsection{Comparison with CVX and SeDuMi}
In this section, we compare the proposed matrix stuffing and ADMM algorithm (implemented in the software package SCS\footnote{https://github.com/cvxgrp/scs.}) with the modeling framework CVX and the interior-point solver SeDuMi.  Consider a network with $L = 100$ $2$-antenna RRHs and
$K = 50$ single-antenna MUs uniformly and independently
distributed in the square region $[-1000, 1000] \times [-1000, 1000]$
meters. We consider a particular network realization for the optimization problem $\mathscr{P}$\footnote{The channel coefficients can be found at http://ihome.ust.hk/$\sim$yshiac/.}.

\begin{table}[!t]
\renewcommand{\arraystretch}{1.3}
\caption{Time in [s] Comparison with CVX and Proposed Matrix Stuffing}
\label{mats_table}
\centering
\begin{tabular}{c|c|c|c|c|c}
{{SINR [dB]}} & {{0}} & 2 & 4 & 6 & 8\\
\hline
CVX & 18.87 & 17.86 & 17.53 & 17.36 & 17.17\\
\hline
Matrix Stuffing & 0.26 & 0.24 & 0.24 & 0.24 & 0.27
\end{tabular}
\end{table} 

\begin{table}[!t]
\renewcommand{\arraystretch}{1.3}
\caption{Time in [s] Comparison with Different Solvers}
\label{algorithm}
\centering
\begin{tabular}{c|c|c|c|c|c}
{{SINR [dB]}} & {{0}} & 2 & 4 & 6 & 8 \\
\hline
SeDuMi& 2668.43 &  2255.29 & 1879.07 & 1518.83 & 1569.06 \\
\hline
SCS & 1.17 & 2.29 & 3.99 & 6.29 & 10.06
\end{tabular}
\label{solver_table}
\end{table}

\begin{table}[!t]
\renewcommand{\arraystretch}{1.3}
\caption{Comparison with the Optimal Values IN [dBm] of Different Algorithms}
\label{algorithm}
\centering
\begin{tabular}{c|c|c|c|c|c}
{{SINR [dB]}} & {{0}} & 2 & 4 & 6 & 8 \\
\hline
CVX+SeDuMi& 33.29 & 35.49 & 37.69 & 39.86 &41.99\\
\hline
Matrix Stuffing+ SCS & 33.28 & 35.49 & 37.69 & 39.86 & 41.99
\end{tabular}
\label{algorithm_table}
\end{table}  

Table {\ref{mats_table}} compares the time consumption of the proposed matrix stuffing and CVX for transforming the original problem $\mathscr{P}$ to the equivalent ADMM-compliant problem $\mathscr{P}_{\textrm{ADMM}}$. This table demonstrates that the time for transformation of proposed matrix stuffing technique is less than one second and can speedup about 200 times compared to the modeling framework CVX.    

 Table {\ref{solver_table}} compares the ADMM method with the interior-point method. With closed-form solutions for the subproblems in the ADMM algorithm and parallelized projection on the cones (\ref{prox}), the ADMM algorithm can speedup in several orders of magnitude over the interior-point method (i.e., about 1000 times from this table).   

Table {\ref{algorithm_table}} presents the optimal values obtained from the modeling framework CVX with the interior-point solver SeDuMi and the proposed large-scale optimization framework. We can see that the proposed framework can provide a solution to modest accuracy with much less time.

\subsection{Minimum Network-Wide Achievable Rate Versus SNR}
\begin{figure}[h]
\centering
\includegraphics[width=1\columnwidth, height=2.5in]{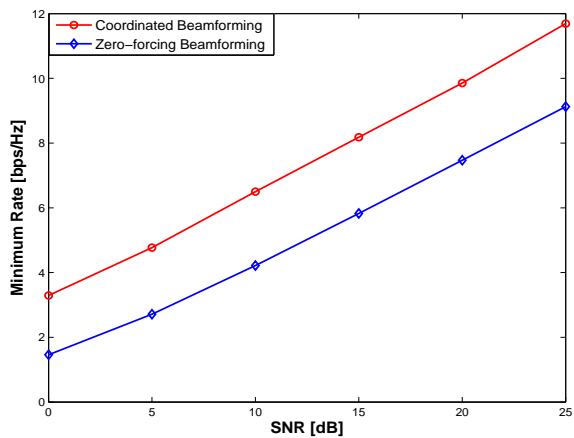}
\caption{The minimum network-wide achievable versus transmit SNR.}
\label{snr}
\end{figure}

Consider a network with $L = 100$ single-antenna RRHs and
$K = 50$ single-antenna MUs uniformly and independently
distributed in the square region $[-5000, 5000] \times [-5000, 5000]$
meters. Fig. {\ref{snr}} demonstrates the minimum network-wide achievable rate versus different SNRs. Each point of the simulation results is averaged over 1000 randomly generated network realizations. From this figure, we can see that the optimal coordinated beamforming can improve the per-user rate by 2.0 {bps/Hz} compared to the zero-forcing beamforming, which is quite an improvement at low to medium SNRs. This motives us to develop the large-scale optimization algorithm to find the optimal coordinated beamformers to improve the spectral efficiency.    

\subsection{Capacity Scaling Law of Large-Scale Networks}
\begin{figure}[!t]
\centering
\includegraphics[width=1\columnwidth, height=2.5in]{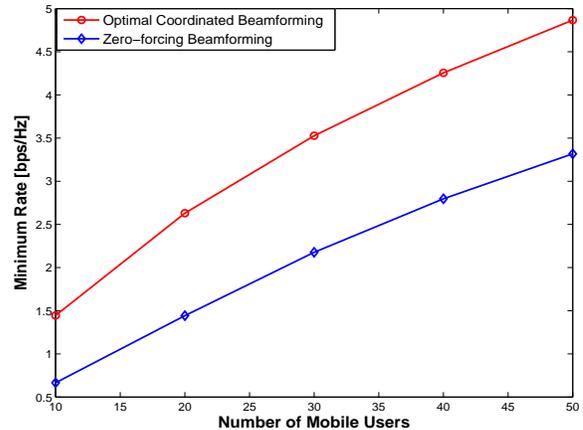}
\caption{The minimum network-wide achievable rate versus user density.}
\label{scaling}
\end{figure}

Consider a network with $L$ 2-antenna RRHs and $K$ single-antenna MUs uniformly and independently distributed in the square region $[-5000, 5000] \times [-5000, 5000]$ meters. The SNR is set as 10 dB. We keep the ratio of $\beta={K/L}=1$ and investigate the rate scaling law of large-scale cooperative networks. Fig. {\ref{scaling}} demonstrates the minimum network-wide achievable
rate versus the user density. Each point of the simulation results is averaged
over 400 randomly generated network realizations. From this figure, we can see that the performance gap between the optimal
coordinated beamforming and the zero-forcing beamforming becomes larger when the network density is higher.
Thus optimal beamforming is required for dense networks. 

\section{Conclusions}
In this paper, we proposed a framework for large-scale optimization for coordinated beamforming  in dense wireless cooperative networks. We first proposed to use the matrix stuffing method to transform the original problem to the equivalent ADMM-compliant problem. The ADMM algorithm was then applied to solve the transformed ADMM-compliant problem. Simulation results have demonstrated the advantage and thus the necessity of the proposed large-scale optimal beamforming algorithm.

\bibliographystyle{ieeetr}
\bibliography{/Users/yuanmingshi/Dropbox/Reference/Reference}

\begin{thebibliography}{10}

\bibitem{Andrews_TIT2013}
A.~Lozano, R.~Heath, and J.~Andrews, ``Fundamental limits of cooperation,''
  {\em IEEE Trans. Inf. Theory}, vol.~59, pp.~5213--5226, Sep. 2013.

\bibitem{Yuanming_arXivOpt13}
Y.~Shi, J.~Zhang, and K.~B. Letaief, ``Optimal stochastic coordinated
  beamforming with compressive {CSI} acquisition for {C}loud-{RAN} (longer
  version with detailed proofs),'' {\em arXiv preprint arXiv:1312.0363}, 2013.

\bibitem{SeDuMi_1999using}
J.~F. Sturm, ``Using {S}e{D}u{M}i 1.02, a {MATLAB} toolbox for optimization
  over symmetric cones,'' {\em Optim. Methods Softw.}, vol.~11, no.~1-4,
  pp.~625--653, 1999.

\bibitem{WeiYu_WC10}
H.~Dahrouj and W.~Yu, ``Coordinated beamforming for the multicell multi-antenna
  wireless system,'' {\em IEEE Trans. Wireless Commun.}, vol.~9,
  pp.~1748--1759, Sep. 2010.

\bibitem{boyd2011distributed}
S.~Boyd, N.~Parikh, E.~Chu, B.~Peleato, and J.~Eckstein, ``Distributed
  optimization and statistical learning via the alternating direction method of
  multipliers,'' {\em Foundations and Trends in Machine Learning}, vol.~3,
  pp.~1--122, Jul. 2011.

\bibitem{shen2012distributed}
C.~Shen, T.-H. Chang, K.-Y. Wang, Z.~Qiu, and C.-Y. Chi, ``Distributed robust
  multicell coordinated beamforming with imperfect {CSI}: an {ADMM} approach,''
  {\em IEEE Trans. Signal Process.}, vol.~60, pp.~2988--3003, Jun. 2012.

\bibitem{boyd_code2013}
E.~Chu, N.~Parikh, A.~Domahidi, and S.~Boyd, ``Code generation for embedded
  second-order cone programming,'' in {\em Control Conference (ECC), 2013
  European}, pp.~1547--1552, Jul. 2013.

\bibitem{cvx}
{CVX Research, Inc.}, ``{CVX}: Matlab software for disciplined convex
  programming, version 2.0 (beta),'' 2013.

\bibitem{boyd2004convex}
S.~P. Boyd and L.~Vandenberghe, {\em Convex optimization}.
\newblock Cambridge university press, 2004.

\bibitem{boyd2008graph}
M.~C. Grant and S.~P. Boyd, ``Graph implementations for nonsmooth convex
  programs,'' in {\em Recent advances in learning and control}, pp.~95--110,
  Springer, 2008.

\bibitem{ye1994nl}
Y.~Ye, M.~J. Todd, and S.~Mizuno, ``An $\mathcal{O}(\sqrt{n}{L})$-iteration
  homogeneous and self-dual linear programming algorithm,'' {\em Math. Oper.
  Res.}, vol.~19, no.~1, pp.~53--67, 1994.

\bibitem{Boyd_arXiv2013}
B.~O'Donoghue, E.~Chu, N.~Parikh, and S.~Boyd, ``Operator splitting for conic
  optimization via homogeneous self-dual embedding,'' {\em arXiv preprint
  arXiv:1312.3039}, 2013.

\bibitem{Boyd_2013proximal}
N.~Parikh and S.~Boyd, ``Proximal algorithms,'' {\em Foundations and trends in
  optimization}, vol.~1, Jan. 2014.

\end{thebibliography}

\end{document}